\documentclass[11pt]{article}
\usepackage{graphicx}

\begin{document}

\title{\textbf{\textsf{On the resolution of cosmic coincidence problem and
phantom crossing with triple interacting fluids}}}
\author{ \textsf{Mubasher Jamil}\footnote{mjamil@camp.edu.pk},\ \
\textsf{Farook Rahaman}\footnote{farook\_rahaman@yahoo.com}
 \\ \\
$^\ast$\small Center for Advanced Mathematics and Physics, National
University of Sciences and Technology,\\ \small Peshawar Road,
Rawalpindi - 46000, Pakistan
\\
$^\dag$\small Department of Mathematics, Jadavpur University,
Kolkata - 700032, India
\\
 }\maketitle
\begin{abstract}
We here investigate a cosmological model in which three fluids
interact with each other involving certain coupling parameters and
energy exchange rates. The motivation of the problem stems from the
puzzling `triple coincidence problem' which naively asks why the
cosmic energy densities of matter, radiation and dark energy are
almost of the same order of magnitude at the present time. In our
model, we determine the conditions under triple interacting fluids
will cross the phantom divide.
\end{abstract}

\textbf{Keywords}: Interacting dark energy; cosmological constant;
cosmic coincidence problem; phantom energy; phantom crossing. \large

\newpage

\section{Introduction}

Despite several successes of the standard big bang cosmology based
on Friedmann-Robertson-Walker (FRW) model, still a series of
problems to be resolved like the horizon problem, flatness problem,
dark matter (or missing mass) problem, structure formation,
topological defects, matter-antimatter asymmetry and the cosmic
coincidence problem etc. Most of these are related with the cosmic
past of the observable universe while the cosmic coincidence problem
has its origin in the recent time since it naively asks why certain
cosmological phenomena are occurring in our presence or in our
times. Recent astrophysical observations give convincing evidence of
an accelerating universe caused by dark energy characterized by the
equation of state (EoS) parameter $\omega\simeq-1$. It is yet
unknown why the present energy density of the dark energy is
approximately equal to that of dust-like matter. It is termed as the
cosmic coincidence problem \cite{stein}. Till now several models
have been proposed in an attempt to solve this problem such as the
`tracker field' \cite{wang}, oscillating dark energy \cite{Noji} and
the variable constants approach \cite{farook}, to name a few. It
appears that the energy density of the radiation component is also
almost equivalent to that of the matter and the dark energy i.e.
$\rho_m\sim\rho_r\sim\rho_\Lambda$ or
$\Omega_m\sim\Omega_r\sim\Omega_\Lambda$, the so-called `cosmic
triple coincidence problem' \cite{hamed}. The question is: why this
happens in the current or in recent times. The history of the
parameter $\omega$ suggests that it is no more a constant but
possesses a parametric form $\omega(z)$, where $z$ is the redshift
parameter. Thus in the past $\omega=1/3$ corresponds to radiation
and then $\omega=0$ for matter. Later it evolved to quintessence
$\omega<-1/3$ to cosmological constant $\omega=-1$. This behavior
suggests that in future, $\omega$ will be super-negative i.e.
$\omega<-1$, which corresponds to the phantom energy. Thus in
totality, we have the transition from $\omega>-1$ to $\omega<-1$ the
so-called `phantom crossing' or `phantom divide' scenario, while we
are observing $\omega=-1$ at the current time \cite{hrvoje}. The
coincidence problem in this context is rephrased as `why is
$\omega=-1$ now?'

In recent years, the usual coincidence problem is addressed by
proposing an exotic interaction between dark energy and matter in
which energy from $\rho_\Lambda$ is diluted or decayed into the
$\rho_m$
\cite{jamil,jamil1,jamil2,marek,barrow,macorra,neto,setare,jamil3,nojiri,winfried,cimento}.
It is recently proposed that if these two components interact then
some energy might dissipate into a third component $\rho_x$ which is
as yet hypothetical \cite{cruz}. The third component can be known
form of matter or an altogether exotic fluid in which case some new
physics will be required to explain the interaction. If we assume
$\rho_x=\rho_r$ then the interaction between three fluids i.e.
matter, radiation and dark energy will be quite interesting. It is
well-known that matter and radiation were decoupled at the time of
emission of cosmic microwave background (CMB) radiation at a
redshift $z\sim 1100$. Thus both matter and radiation are almost
non-interacting components but it can be anticipated that these two
components do supposedly interact with the dark energy. Thus if dark
energy and matter interact, the energy dissipated in the interaction
is assumed to transfer to the radiation component and vice versa for
the radiation and the dark energy interaction. This dynamic
interaction than hugely alters the effective equation of state of
the three interacting fluids. Our analysis in this paper suggests
that effective EoS of the three fluids will be interlinked to each
other considerably and also constrained by coupling parameters.
Moreover, the interaction naturally leads to the phantom crossing
scenario.

We here present a model in which three cosmic fluids (radiation,
matter and dark energy) interact with each other with equal degree
of freedom. Hence there are three coupling parameters involved which
can take arbitrary positive or negative values not all zero
simultaneously. The positive and negative values correspond to back
and forth nature of energy exchange between the fluids. We emphasis
here that the coupling parameters in our model are only constrained
by the choices of effective EoS of the fluids. Thus in our model,
exact EoS of the fluids like in the non-interacting FRW model is not
possible. As the exact EoS for the dark energy is unknown, its
interaction with other fluids creates ambiguities in the
determination of the exact EoS of the other fluids. Hence we also
stress that precise values of the EoS's can be deduced only
empirically from the phenomenology of the interacting fluids.

\section{The model of triple interacting fluids}
We start by assuming the background to be spatially flat,
homogeneous and isotropic FRW spacetime
\begin{equation}
ds^2=-dt^2+a^2(t)\left[dr^2+ r^2(d\theta^2+\sin^2\theta
d\phi^2)\right],
\end{equation}
where $a(t)$ is the scale factor. We consider three fluids having
the equations of state (EoS) $p_i=\omega_i\rho_i$, $i=1,2,3$ where
$p_i$ and $\rho_i$ are the corresponding pressures and the energy
densities of the fluids, respectively. Also $\omega_i$ are the
dimensionless EoS parameters. For the sake of simplicity, we assume
the three fluids to be perfect fluid-like since in general,
interaction between fluids might lead to local inhomogeneities,
which are ignored in the present paper. The equations governing the
interaction of three fluids are expressed as
\begin{eqnarray}
\dot{\rho}_1+3H(1+\omega_1)\rho_1&=&Q_3-Q_2,\\
\dot{\rho}_2+3H(1+\omega_2)\rho_2&=&Q_1-Q_3,\\
\dot{\rho}_3+3H(1+\omega_3)\rho_3&=&Q_2-Q_1.
\end{eqnarray}
Here  $Q_i$ are the energy exchange (or dissipative) terms to be put
ad hoc in the above equations. The explicit form of $Q_i$ should be
determined only from the phenomenological and empirical results.
However, from dimensional considerations, the quantity $Q_i$ should
have dimensions of density into the time inverse. Choosing the later
one to be Hubble parameter, we notice that $Q_i$ can be of the
following forms: $Q_i\simeq H\rho_i$. This approximation can be
saturated to equality by inserting a dimensionless parameter (say
$\lambda_i$). Thus we can write \cite{mohseni} (see \cite{wei} for
more exotic expressions for $Q_i$)
\begin{equation}
Q_i=\lambda_iH\rho_{io}a^{-3(1+\omega_i)}, \ \ i=1,2,3.
\end{equation}
In the last expression, $\lambda_i$ are the coupling constants which
can take positive or negative values to yield two-sided energy
exchange rather than one-sided. Also Eqs. (2) to (5) show that this
a coupled system of three differential equations which needs to be
solved. Further $\rho_{io}$ are the constant energy densities at
some reference time $t=t_o$. Also $H\equiv\dot{a}/a$ is Hubble
parameter which determines the rate of expansion of the universe.
Sum of Eqs. (2) to (4) yield the combined energy conservation
\begin{equation}
\sum_{i=1}^3[\dot{\rho_i}+3H(1+\omega_i)\rho_i]=0,
\end{equation}
Also the energy conservation for the individual component (for the
case of non-interacting fluids) yields
\begin{equation}
\rho_i^\prime=\rho_{io}a^{-3(1+\omega_i)}, \ \ i=1,2,3.
\end{equation}
Here $\rho_{io}$ are integration constants. Combining Eqs. (2), (3)
and (4), we arrive at the density evolution of the interacting
fluids as
\begin{eqnarray}
\rho_1&=&C_1a^{-3(1+\omega_1)}+\frac{\lambda_2\rho_{2o}}{3}\frac{a^{-3(1+\omega_2)}}{\omega_2-\omega_1}+\frac{\lambda_3\rho_{3o}}{3}\frac{a^{-3(1+\omega_3)}}{\omega_1-\omega_3},\\
\rho_2&=&C_2a^{-3(1+\omega_2)}+\frac{\lambda_3\rho_{3o}}{3}\frac{a^{-3(1+\omega_3)}}{\omega_3-\omega_2}+\frac{\lambda_1\rho_{1o}}{3}\frac{a^{-3(1+\omega_1)}}{\omega_2-\omega_1},\\
\rho_3&=&C_3a^{-3(1+\omega_3)}+\frac{\lambda_1\rho_{1o}}{3}\frac{a^{-3(1+\omega_1)}}{\omega_1-\omega_3}+\frac{\lambda_2\rho_{2o}}{3}\frac{a^{-3(1+\omega_2)}}{\omega_3-\omega_2}.
\end{eqnarray}
Here $C_i$'s are constants of integration. Now addition of Eqs. (8),
(9) and (10) results in
\begin{equation}
\rho=\rho_1^\prime X_1+\rho_2^\prime X_2+\rho_3^\prime X_3,
\end{equation}
where
\begin{eqnarray}
X_1&=&
1+\frac{\lambda_1}{3}\frac{\omega_2-\omega_3}{(\omega_1-\omega_3)(\omega_2-\omega_1)},\\
X_2&=&
1+\frac{\lambda_2}{3}\frac{\omega_3-\omega_1}{(\omega_2-\omega_1)(\omega_3-\omega_2)},\\
X_3&=&
1+\frac{\lambda_3}{3}\frac{\omega_1-\omega_2}{(\omega_1-\omega_3)(\omega_3-\omega_2)}.
\end{eqnarray}
Here we have assumed $C_i=\rho_{io}$ and
$\rho\equiv\rho_1+\rho_2+\rho_3$ is the total energy density of the
interacting fluids. Note that the case of non-interacting fluids is
obtained by choosing $\lambda_1=\lambda_2=\lambda_3=0$ which yield
$X_1=X_2=X_3=1$ i.e.
$\rho=\rho_1^\prime+\rho_2^\prime+\rho_3^\prime$. Using the first
FRW equation
\begin{equation}
H^2=\frac{8\pi G}{3}\rho.
\end{equation}
Differentiating Eq. (7) w.r.t $t$, we obtain
\begin{equation}
\dot{\rho}^\prime_i=-3(1+\omega_i)H\rho_{io}a^{-3(1+\omega_i)}, \ \
i=1,2,3.
\end{equation}
Differentiating Eq. (15) w.r.t $t$ and then using Eqs. (11) and
(16), we find
\begin{eqnarray}
\dot{H}&=&-H\sqrt{24\pi
G}\sum_{i=1}^3[\rho_i^\prime(1+\omega_i)X_i]\left(\sum_{i=1}^3\rho_i^\prime
X_i\right)^{-1/2}.
\end{eqnarray}
\subsection{Role of parameter $\dot H$}
The parameter $\dot H$ is significant as its possible signature
governs the dynamics of the universe \cite{brevik}. The slowing down
in the expansion takes place when
\begin{equation}
\omega_1>-1,\ \
\lambda_1>\frac{3(\omega_1-\omega_3)(\omega_2-\omega_1)}{\omega_3-\omega_2},
\end{equation}
\begin{equation}
\omega_2>-1,\ \
\lambda_2>\frac{3(\omega_2-\omega_1)(\omega_3-\omega_2)}{\omega_1-\omega_3},
\end{equation}
\begin{equation}
\omega_3>-1,\ \
\lambda_3>\frac{3(\omega_1-\omega_3)(\omega_3-\omega_2)}{\omega_2-\omega_1}.
\end{equation}
Thus $\dot{H}$ will be negative when all the $\omega_i>-1$. This
result corresponds to the quintessence dominated universe.

Note that the vanishing $\dot{H}$ in Eq. (17) will yield a de Sitter
universe or a cosmological constant dominated universe i.e.
\begin{equation}
\dot{H}=0 \Longrightarrow \omega_1=\omega_2=\omega_3=-1.
\end{equation}
Moreover $\dot{H}>0$ corresponds to an accelerating universe. This
situation arises in our model when
\begin{eqnarray}
(1+\omega_1)X_1&<&0,\\
(1+\omega_2)X_2&<&0,\\
(1+\omega_3)X_3&<&0.
\end{eqnarray}
The above Eqs. (22), (23) and (24) yield respectively
\begin{equation}
\omega_1<-1,\ \
\lambda_1>\frac{3(\omega_1-\omega_3)(\omega_2-\omega_1)}{\omega_3-\omega_2},
\end{equation}
\begin{equation}
\omega_2<-1,\ \
\lambda_2>\frac{3(\omega_2-\omega_1)(\omega_3-\omega_2)}{\omega_1-\omega_3},
\end{equation}
\begin{equation}
\omega_3<-1,\ \
\lambda_3>\frac{3(\omega_1-\omega_3)(\omega_3-\omega_2)}{\omega_2-\omega_1}.
\end{equation}
Thus $\dot{H}$ will be positive when all the $\omega_i<-1$. This
result corresponds to the phantom energy dominated universe. Note
that the coupling parameters $\lambda_i$ have to be positive both
for the decelerating and the accelerating universe. This result
turns out to be consistent with \cite{feng1} that coupling
parameters cannot be negative to avoid violation of second law of
thermodynamics. Moreover the same investigation shows that small
positive values for the coupling parameters are motivated from the
empirical results. Therefore the universe evolves from the earlier
quintessence to cosmological constant and then later to the phantom
energy dominated universe. Also the conditions (18)-(20), and the
analysis of this section determines the super-acceleration and not
the acceleration.

\subsection{Behavior of deceleration parameter}

The deceleration parameter $q\sim\dot{H} + H^2$ = $ -8 \pi G
\sum_{i=1}^3[\rho_i^\prime(\frac{2}{3}+\omega_i)X_i]$. The parameter
$q$ is significant as its possible signature governs the dynamics of
the universe. For instance $q<0$ represents the acceleration phase
of the expanding universe. This slowing down in the expansion takes
place when
\begin{equation}
\omega_1>-\frac{2}{3},\ \
\lambda_1>\frac{3(\omega_1-\omega_3)(\omega_2-\omega_1)}{\omega_3-\omega_2},
\end{equation}
\begin{equation}
\omega_2>-\frac{2}{3},\ \
\lambda_2>\frac{3(\omega_2-\omega_1)(\omega_3-\omega_2)}{\omega_1-\omega_3},
\end{equation}
\begin{equation}
\omega_3>-\frac{2}{3},\ \
\lambda_3>\frac{3(\omega_1-\omega_3)(\omega_3-\omega_2)}{\omega_2-\omega_1}.
\end{equation}
Thus $q$ will be negative when all the $\omega_i>-\frac{2}{3}$. This
result corresponds to the quintessence dominated universe.

Note that the vanishing $q$ in Eq. (17) will yield the following
state equations
\begin{equation}
q=0 \Longrightarrow \omega_1=\omega_2=\omega_3=-\frac{2}{3}.
\end{equation}
Moreover $q>0$ corresponds to an decelerating universe. This
situation arises in our model when
\begin{eqnarray}
(\frac{2}{3}+\omega_1)X_1&<&0,\\
(\frac{2}{3}+\omega_2)X_2&<&0,\\
(\frac{2}{3}+\omega_3)X_3&<&0.
\end{eqnarray}
The above Eqs. (32), (33) and (34) yield respectively
\begin{equation}
\omega_1<-\frac{2}{3},\ \
\lambda_1>\frac{3(\omega_1-\omega_3)(\omega_2-\omega_1)}{\omega_3-\omega_2},
\end{equation}
\begin{equation}
\omega_2<-\frac{2}{3},\ \
\lambda_2>\frac{3(\omega_2-\omega_1)(\omega_3-\omega_2)}{\omega_1-\omega_3},
\end{equation}
\begin{equation}
\omega_3<-\frac{2}{3},\ \
\lambda_3>\frac{3(\omega_1-\omega_3)(\omega_3-\omega_2)}{\omega_2-\omega_1}.
\end{equation}
Thus $q$ will be negative when all the $\omega_i<-\frac{2}{3}$. The
strength of the interaction is determined by the coupling parameters
$\lambda_i$. These coupling constants  are expressed in terms of the
equation of state parameters $\omega_i$. We here stress that the
exact nature of the interaction is largely unknown i.e. the
mediating particles of the interaction are not yet identified. Any
interacting dark energy model should, in principle, be motivated
from the particle physics or the corresponding phenomenology,
however there are as yet no sound theoretical models which could
identify the particle interactions. There are some arguments that
the phantom-like dark energy can decay into at least one ordinary
particle and some other phantom-like particles \cite{carroll}.

\section{Stability analysis}
To perform stability analysis, we write Eqs. (2-5) as
\begin{eqnarray}
\dot\rho_1+3H(1+\omega_1)\rho_1&=&3H\lambda_1(\rho_3-\rho_2),\\
\dot\rho_2+3H(1+\omega_2)\rho_2&=&3H\lambda_2(\rho_1-\rho_3),\\
\dot\rho_3+3H(1+\omega_3)\rho_3&=&3H\lambda_3(\rho_2-\rho_1).
\end{eqnarray}
Here we have used the constraint equations on the coupling
parameters by: $\lambda_3-\lambda_2=\lambda_1$,
$\lambda_1-\lambda_3=\lambda_2$ and $\lambda_2-\lambda_1=\lambda_3$.
Since the above system is autonomous i.e. there is no explicit time
dependent term in the equations, we can analyze the system by first
finding its critical points and later on, checking the stability of
the system about those points. A critical point (also called fixed
point) is the one that satisfies the dynamical system (like Eqs.
2-5) when equated to zero. Similarly, a critical point becomes an
attractor if the solution of the system converges to that point for
large values of the time parameter. Such kind of analysis is
generally performed to check whether the dynamical system possesses
any solution which is stable against small perturbations
\cite{prl,ijmpd,jcap,plb,prd,jcap1}.

Further, Defining the dimensionless density parameters as
\begin{eqnarray}
u_1&=&\frac{\rho_1}{\rho_{cr}}=\Omega_1,\\
u_2&=&\frac{\rho_2}{\rho_{cr}}=\Omega_2,\\
u_3&=&\frac{\rho_3}{\rho_{cr}}=\Omega_3.
\end{eqnarray}
Using Eqs. (41-43), we can rewrite Eqs (38-40) as
\begin{eqnarray}
\frac{du_1}{dx}&=& 3u_1\{(1+\omega_2)u_2+(1+\omega_3)u_3\}+3\lambda_1(u_3-u_2) ,\\
\frac{du_2}{dx}&=& 3u_2\{(1+\omega_1)u_1+(1+\omega_3)u_3\}+3\lambda_2(u_1-u_3) ,\\
\frac{du_3}{dx}&=&
3u_3\{(1+\omega_1)u_1+(1+\omega_2)u_2\}+3\lambda_3(u_2-u_1).
\end{eqnarray}
Above the time derivative has been replaced with the differentiation
with respect to a new parameter $x=\ln{a}$. This parameter
corresponds to the number of e-foldings which is convenient to use
for the dynamics of dark energy. By equating above three equations
to zero, we obtain the critical points (see below). To find whether
the system approaches to any critical point, we check the stability
about these points, by producing small perturbations $\delta u_1$,
$\delta u_2$ and $\delta u_3$ around the critical point
$(u_{1_c},u_{2_c},u_{3_c})$ i.e.
\begin{equation}
u_1=u_{1_c}+\delta u_1, \ \ u_2=u_{2_c}+\delta u_2,\ \
u_3=u_{3_c}+\delta u_3.
\end{equation}
Now in order to linearize our dynamical system, we take the $\delta$
variation of the above equations (44-46) and evaluate them at the
critical points $\{(u_{1_{c_j}}, u_{2_{c_j}},u_{3_{c_j}})\}$, where
$j$ is an indexing parameter:
\begin{eqnarray}
\frac{d\delta u_{1}}{dx}&=&3\delta u_{1}\{
(1+\omega_2)u_{2_{c_j}}+(1+\omega_3)u_{3_{c_j}} \}+3\delta u_{2} \{
u_{1_{c_j}}(1+\omega_2)-\lambda_1 \}\nonumber\\
&\;&+3\delta u_{3} \{ u_{1_{c_j}}(1+\omega_3)+\lambda_1 \},\\
\frac{d\delta u_{2}}{dx}&=&3\delta u_{2}\{
(1+\omega_1)u_{1_{c_j}}+(1+\omega_3)u_{3_{c_j}} \}+3\delta u_{3} \{
u_{2_{c_j}}(1+\omega_3)-\lambda_2 \}\nonumber\\
&\;&+3\delta u_{1} \{ u_{2_{c_j}}(1+\omega_1)+\lambda_2 \},\\
\frac{d\delta u_{3}}{dx}&=&3\delta u_{3}\{
(1+\omega_2)u_{2_{c_j}}+(1+\omega_1)u_{1_{c_j}} \}+3\delta u_{1} \{
u_{3_{c_j}}(1+\omega_1)-\lambda_3\}\nonumber\\
&\;&+3\delta u_{2} \{ u_{3_{c_j}}(1+\omega_2)+\lambda_3 \}.
\end{eqnarray}
We then next construct a matrix consisting of the coefficients of
the perturbation parameters and compute the eigenvalues of that
matrix. The stability of the critical points depends on the nature
of the corresponding eigenvalues and it leads to three specific
cases: if the real parts of all the eigenvalues are negative, then
the following critical point is a stable node; if the real parts of
the eigenvalues are all positive, then that critical point is an
unstable node, in all other cases, the critical points will be
saddle points. In our analysis, we shall be interested in only those
critical points that yield stable nodes.

In order to obtain some physical interpretation of our results, we
assign dust (or matter), radiation and dark energy with subscripts
$1,2$ and $3$ respectively. Thus the corresponding EoS parameters
take values $\omega_1=0$, $\omega_2=1/3$ and $\omega_3=-1/3$ or $-1$
depending whether it is quintessence or cosmological constant. From
here onwards, we divide our dynamical system (44-46) into two parts:
one for $\omega_3=-1$ and other for $\omega_3=-1/3$.

It is not possible to obtain the critical points of the full system
(44-46). The only way out is by choosing at least one parameter
$\lambda_i$ to be zero. In the next sections, we shall use only two
coupling parameters while taking the third one to be zero. For
instance, taking $\lambda_1=0$ implies that radiation and dark
energy are mutually non-interacting while there are interactions
between dark energy and matter and between matter and radiation.
Similar interpretations can be made when choosing either $\lambda_2$
or $\lambda_3$ to be zero.

\subsection{Critical points for $\lambda_1=0$}
Note that the following critical points are obtained by setting
$\lambda_1=0.$
\subsubsection{Quintessence $\omega_3=-1/3$}
The dynamical system yields the following critical points
$\{(u_{1_{c_j}}, u_{2_{c_j}},u_{3_{c_j}}), j=1,2,3,4\}$.
\begin{eqnarray}
u_{1_{c_1}}&=&0,\\
u_{2_{c_1}}&=&\frac{3}{2}\lambda_2,\\
u_{3_{c_1}}&=&-\frac{3}{4}\lambda_3,\\
u_{1_{c_2}}&=&\frac{1}{48\lambda_2-24\lambda_3}\Big(-8\lambda_2^2+\lambda_3\Big( \lambda_3-\sqrt{16\lambda_2^2+568\lambda_2\lambda_3+\lambda_3^2} \Big)\nonumber\\
&\;&-2\lambda_2\Big(  -35\lambda_3+\sqrt{16\lambda_2^2+568\lambda_2\lambda_3+\lambda_3^2}   \Big)\Big),\\
u_{2_{c_2}}&=&\frac{1}{32}\Big( 4\lambda_2-\lambda_3+\sqrt{16\lambda_2^2+568\lambda_2\lambda_3+\lambda_3^2}  \Big),\\
u_{3_{c_2}}&=&\frac{1}{16}\Big(
-4\lambda_2+\lambda_3-\sqrt{16\lambda_2^2+568\lambda_2\lambda_3+\lambda_3^2}
\Big),\\
u_{1_{c_3}}&=&\frac{1}{48\lambda_2-24\lambda_3}\Big( -8\lambda_2^2+\lambda_3\Big( \lambda_3+\sqrt{16\lambda_2^2+568\lambda_2\lambda_3+\lambda_3^2} \Big)\nonumber\\
&\;&+2\lambda_2\Big(  -35\lambda_3+\sqrt{16\lambda_2^2+568\lambda_2\lambda_3+\lambda_3^2}   \Big) \Big),\\
u_{2_{c_3}}&=&\frac{1}{32}\Big( 4\lambda_2-\lambda_3-\sqrt{16\lambda_2^2+568\lambda_2\lambda_3+\lambda_3^2}  \Big),\\
u_{3_{c_3}}&=&\frac{1}{16}\Big(
-4\lambda_2+\lambda_3+\sqrt{16\lambda_2^2+568\lambda_2\lambda_3+\lambda_3^2}
\Big),\\
u_{4_{c_1}}&=&0,\\
u_{4_{c_2}}&=&0,\\
u_{4_{c_3}}&=&0.
\end{eqnarray}
Now we substitute above critical points in Eqs. (38-40) and obtain
eigenvalues. Since analytical expressions for the eigenvalues are
not possible, so we shall resort to estimate the numerical values of
eigenvalues for specific choices of coupling parameters. Fixing
$\lambda_2=0.5$ and $\lambda_3=0.7$ (both positive), $\lambda_2=0.5$
and $\lambda_3=-0.7$ (one positive) and $\lambda_2=-0.5$ and
$\lambda_3=-0.7$ (both negative) no stable node is obtained from the
above critical points. However, for $\lambda_2=-0.5$ and
$\lambda_3=0.7$, the first critical point $\{(u_{1_{c_1}},
u_{2_{c_1}},u_{3_{c_1}})\}$ becomes a stable node.

\subsubsection{Cosmological constant $\omega_3=-1$}
\begin{eqnarray}
u_{1_{c_1}}&=&\lambda_3,\\
u_{2_{c_1}}&=&0,\\
u_{3_{c_1}}&=&\lambda_3,\\
u_{1_{c_2}}&=&0,\\
u_{2_{c_2}}&=&0,\\
u_{3_{c_2}}&=&0.
\end{eqnarray}
For all the choices of $\lambda_2$ and $\lambda_3$ as adopted in the
previous subsection, no critical point arises as a stable node.
\subsection{Critical points for $\lambda_2=0$}
Setting $\lambda_2=0$, we obtain the following critical points:
\subsubsection{Quintessence $\omega_3=-1/3$}

\begin{eqnarray}
u_{1_{c_1}}&=&-\frac{1}{2}\lambda_1,\\
u_{2_{c_1}}&=&0,\\
u_{3_{c_1}}&=&\lambda_3,\\
u_{1_{c_2}}&=&\frac{1}{48}\Big( -9\lambda_1-2\lambda_3+\sqrt{81\lambda_1^2+1476\lambda_1\lambda_3+4\lambda_3^2}   \Big),\\
u_{2_{c_2}}&=&\frac{1}{64(3\lambda_1-2\lambda_3)}\Big(  27\lambda_1^2-2\lambda_3\Big( -2\lambda_3+\sqrt{81\lambda_1^2+1476\lambda_1\lambda_3+4\lambda_3^2} \Big)\nonumber\\
&\;&-3\lambda_1\Big( 72\lambda_3+ \sqrt{81\lambda_1^2+1476\lambda_1\lambda_3+4\lambda_3^2}  \Big) \Big),\\
u_{3_{c_2}}&=&\frac{1}{32}\Big( 9\lambda_1+2\lambda_3-\sqrt{81\lambda_1^2+1476\lambda_1\lambda_3+4\lambda_3^2}   \Big),\\
u_{1_{c_3}}&=&\frac{1}{48}\Big( -9\lambda_1-2\lambda_3-\sqrt{81\lambda_1^2+1476\lambda_1\lambda_3+4\lambda_3^2}   \Big),\\
u_{2_{c_3}}&=&\frac{1}{64(3\lambda_1-2\lambda_3)}\Big(  27\lambda_1^2+3\lambda_1\Big( -72\lambda_3+\sqrt{81\lambda_1^2+1476\lambda_1\lambda_3+4\lambda_3^2} \Big)\nonumber\\
&\;&+2\lambda_3\Big( 2\lambda_3+ \sqrt{81\lambda_1^2+1476\lambda_1\lambda_3+4\lambda_3^2}  \Big) \Big),\\
u_{3_{c_3}}&=&\frac{1}{32}\Big(
9\lambda_1+2\lambda_3+\sqrt{81\lambda_1^2+1476\lambda_1\lambda_3+4\lambda_3^2}
\Big),\\
u_{1_{c_4}}&=&0,\\
u_{2_{c_4}}&=&0,\\
u_{3_{c_4}}&=&0.
\end{eqnarray}
For $\lambda_1=0.5$ and $\lambda_3=-0.7$, the stable node arises at
second critical point $\{(u_{1_{c_2}}, u_{2_{c_2}},u_{3_{c_2}})\}$.
For other values of $\lambda_1$ and $\lambda_3$, no other stable
node arises.
\subsubsection{Cosmological constant $\omega_3=-1$}
\begin{eqnarray}
u_{1_{c_1}}&=&0,\\
u_{2_{c_1}}&=&-\frac{3}{4}\lambda_3,\\
u_{3_{c_1}}&=&-\frac{3}{4}\lambda_3,\\
u_{1_{c_2}}&=&0,\\
u_{2_{c_2}}&=&0,\\
u_{3_{c_2}}&=&0.
\end{eqnarray}
Here for $\lambda_1=0.5$ and $\lambda_3=-0.7$, the stable node
arises at second critical point $\{(u_{1_{c_2}},
u_{2_{c_2}},u_{3_{c_2}})\}$.
\subsection{Critical points for $\lambda_3=0$}
Now we perform similar analysis for $\lambda_3=0$, we get the
critical points
\subsubsection{Quintessence $\omega_3=-1/3$}
\begin{eqnarray}
u_{1_{c_1}}&=&\frac{3}{4}\lambda_1,\\
u_{2_{c_1}}&=&-\lambda_2,\\
u_{3_{c_1}}&=&0,\\
u_{1_{c_2}}&=&\frac{1}{48}\Big( -9\lambda_1-8\lambda_2+\sqrt{81\lambda_1^2+4176\lambda_1\lambda_2+64\lambda_2^2}   \Big),\\
u_{2_{c_2}}&=&\frac{1}{64}\Big( 9\lambda_1+8\lambda_2-\sqrt{81\lambda_1^2+4176\lambda_1\lambda_2+64\lambda_2^2}   \Big),\\
u_{3_{c_2}}&=& \frac{1}{96\lambda_1-128\lambda_2}\Big(  27\lambda_1^2+396\lambda_1\lambda_2+32\lambda_2^2-3\lambda_1\sqrt{81\lambda_1^2+4176\lambda_1\lambda_2+64\lambda_2^2}\nonumber\\&\;&-4\lambda_2\sqrt{81\lambda_1^2+4176\lambda_1\lambda_2+64\lambda_2^2}  \Big),\\
u_{1_{c_3}}&=&\frac{1}{48}\Big( -9\lambda_1-8\lambda_2-\sqrt{81\lambda_1^2+4176\lambda_1\lambda_2+64\lambda_2^2}   \Big),\\
u_{2_{c_3}}&=&\frac{1}{64}\Big( 9\lambda_1+8\lambda_2+\sqrt{81\lambda_1^2+4176\lambda_1\lambda_2+64\lambda_2^2}   \Big),\\
u_{3_{c_3}}&=&\frac{1}{96\lambda_1-128\lambda_2}\Big(
27\lambda_1^2+396\lambda_1\lambda_2+32\lambda_2^2+3\lambda_1\sqrt{81\lambda_1^2+4176\lambda_1\lambda_2+64\lambda_2^2}\nonumber\\&\;&+4\lambda_2\sqrt{81\lambda_1^2+4176\lambda_1\lambda_2+64\lambda_2^2}
\Big),\\
u_{1_{c_4}}&=&0,\\
u_{2_{c_4}}&=&0,\\
u_{3_{c_4}}&=&0.
\end{eqnarray}
Among the above critical points, the only stable node is produced
for the first one for values $\lambda_1=-0.5$ and $\lambda_2=0.7$.
\subsubsection{Cosmological constant $\omega_3=-1$}
\begin{eqnarray}
u_{1_{c_1}}&=&\frac{3}{4}\lambda_1,\\
u_{2_{c_1}}&=&-\lambda_2,\\
u_{3_{c_1}}&=&0,\\
u_{1_{c_2}}&=&\frac{7\lambda_1\lambda_2}{3\lambda_1+4\lambda_2},\\
u_{2_{c_2}}&=&-\frac{21\lambda_1\lambda_2}{4(3\lambda_1+4\lambda_2)},\\
u_{3_{c_2}}&=&\frac{7\lambda_1\lambda_2(-9\lambda_1+16\lambda_2)}{4(3\lambda_1+4\lambda_2)^2},\\
u_{1_{c_3}}&=&0,\\
u_{2_{c_3}}&=&0,\\
u_{3_{c_3}}&=&0.
\end{eqnarray}
Among the above critical points, the only stable node is produced
for the first one for values $\lambda_1=-0.5$ and $\lambda_2=0.7$.

In figures 1-6, we provide the schematic representation of triple
interacting fluids, taking two at a time, with various choices of
the coupling parameters and initial conditions. The initial
conditions are chosen arbitrarily but these must satisfy the
constraint $u_1(0)+u_2(0)+u_3(0)=1$ (the Friedmann equation). In
figures 1 and 4 , we assume first and second interacting fluids
whereas in figures 2 and 5 and 3 and 6, we assume second and third
and first and third interacting fluids respectively.  The three
solutions $u_1$, $u_2$ and $u_3$ always satisfy the constraint
$u_1(0)+u_2(0)+u_3(0)=1$.

\section{Conclusion and discussion}

In this paper, we have attempted to resolve the cosmic triple
coincidence problem which naively asks why the energy densities of
the three major ingredients of the cosmic composition namely matter,
radiation and dark energy are of same order at current time.
Rephrasing, why $\omega$ has evolved to $-1$ in recent times. We
here point out that the EoS $p=\omega\rho$ used in this paper is
not, in general, a true EoS for any cosmic fluid. Rather it is a
phenomenological relationship suitable for the configuration. The
actual EoS may not be that simple and may have dependencies on
various other cosmological parameters like redshift, time, Hubble
parameter and its derivatives etc \cite{jamil1}. However to a first
order approximation, it may be taken for such analysis. Further, the
standard non-interacting FRW model cannot resolve the coincidence
problem since it predicts a hierarchial system in which radiation
density decreases faster compared to matter, while density of dark
energy remains either constant (if it is cosmological constant) or
increases (if it is phantom energy). Therefore it contradicts with
the observations where all the three components have equivalent
densities. Model of interacting dark energy, which is a modification
of the FRW model, has enormous potential to explain this
cosmological conundrum. It naturally predicts that if the cosmic
fluids interact with each other, it leads to a scenario compatible
with the observations \cite{campo}.

We have found that in the interacting fluid model, the three fluids
can achieve super-negative equation of state. If matter and
radiation are among the components then they will induce negative
pressure along with the dark energy to produce accelerated
expansion. This result has earlier been shown for the interacting
Chaplygin gas model in \cite{jamil}.

\pagebreak

\begin{figure}
\includegraphics[scale=.7]{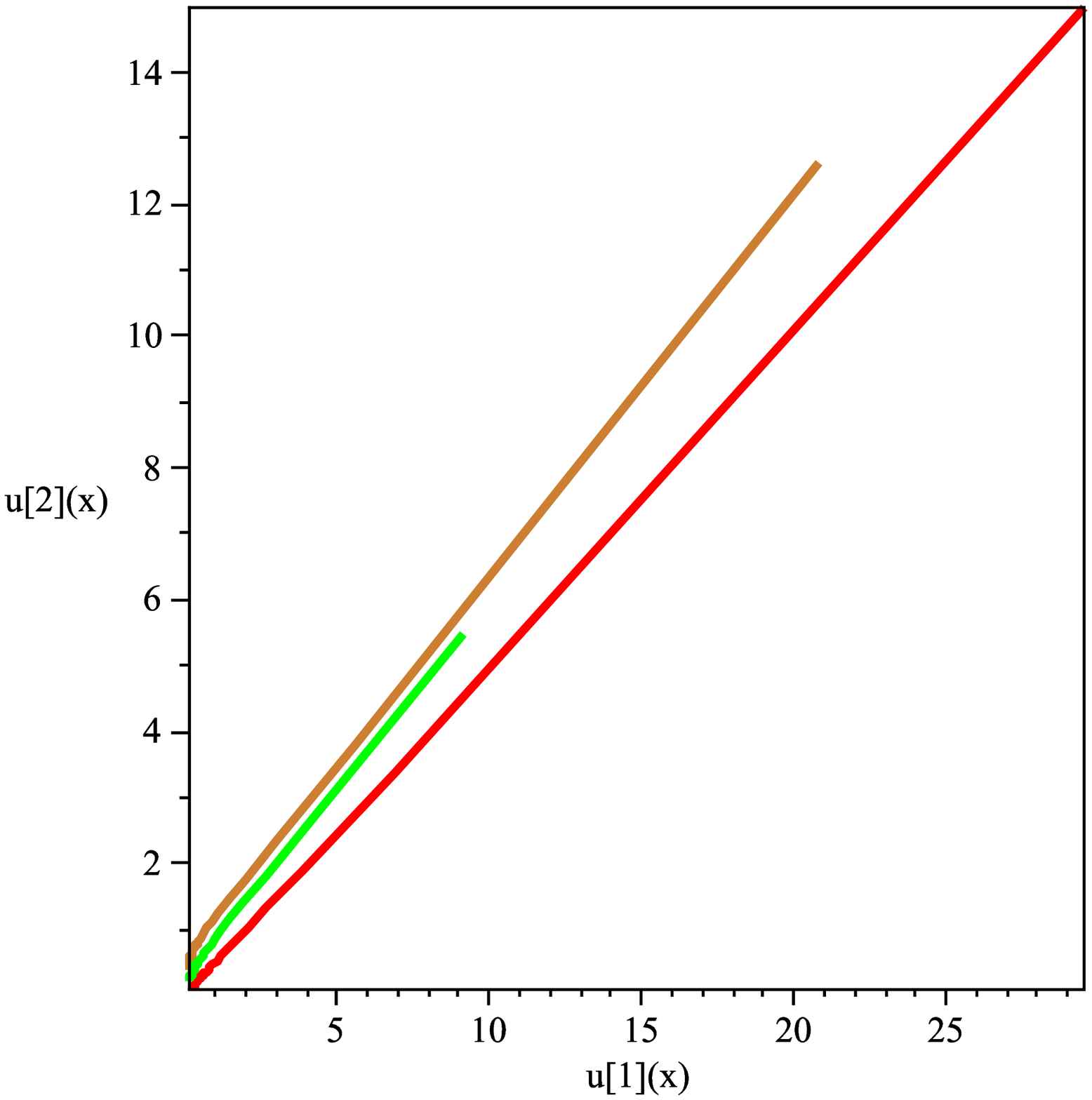}\\
\caption{The phase diagram of the triple interacting fluid model
with the choice of couplings [$\lambda_1=0.2,
\lambda_2=0.04,\lambda_3=0.06$], $x=-1,..,1$ and EoS parameters [
$\omega_1=0,\omega_2=1/3,\omega_3=-1$]. Chosen scene is between
$u_1$ and $u_2$. The curves correspond to the initial conditions
$u_1(0)=0.2, u_2(0)=0.6, u_3(0)=0.2$ (brown); $u_1(0)=0.4,
u_2(0)=0.2, u_3(0)=0.4$ (red); $u_1(0)=0.3, u_2(0)=0.4, u_3(0)=0.3$
(green).}
\end{figure}
\newpage
\begin{figure}
\includegraphics[scale=.7]{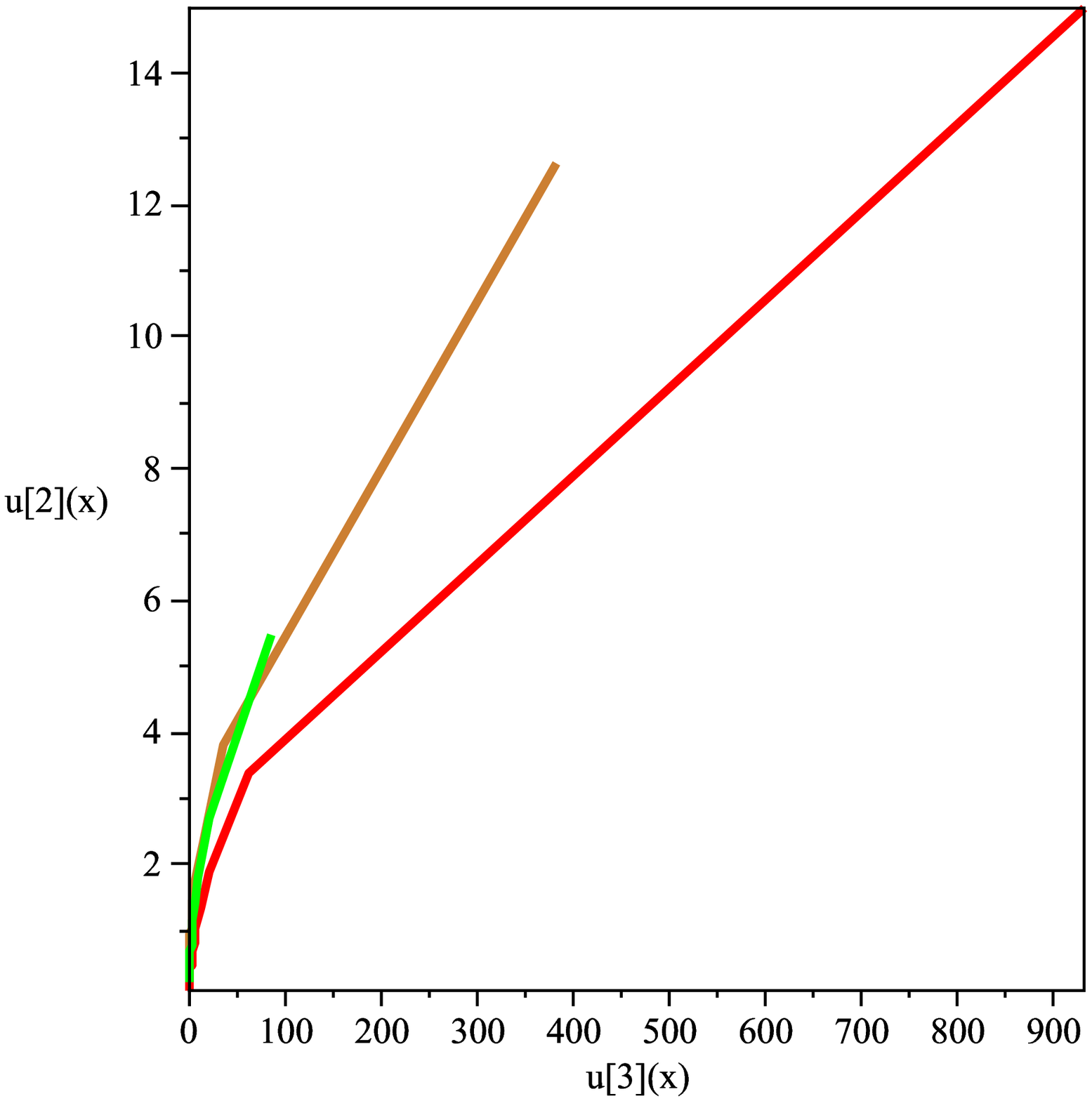}\\
\caption{The phase diagram of the triple interacting fluid model
with the choice of couplings [$\lambda_1=0.2,
\lambda_2=0.04,\lambda_3=0.06$], $x=-1,..,1$ and EoS parameters [
$\omega_1=0,\omega_2=1/3,\omega_3=-1$]. Chosen scene is between
$u_2$ and $u_3$. The curves correspond to the initial conditions as
taken in Fig.1.}
\end{figure}
\newpage
\begin{figure}
\includegraphics[scale=.7]{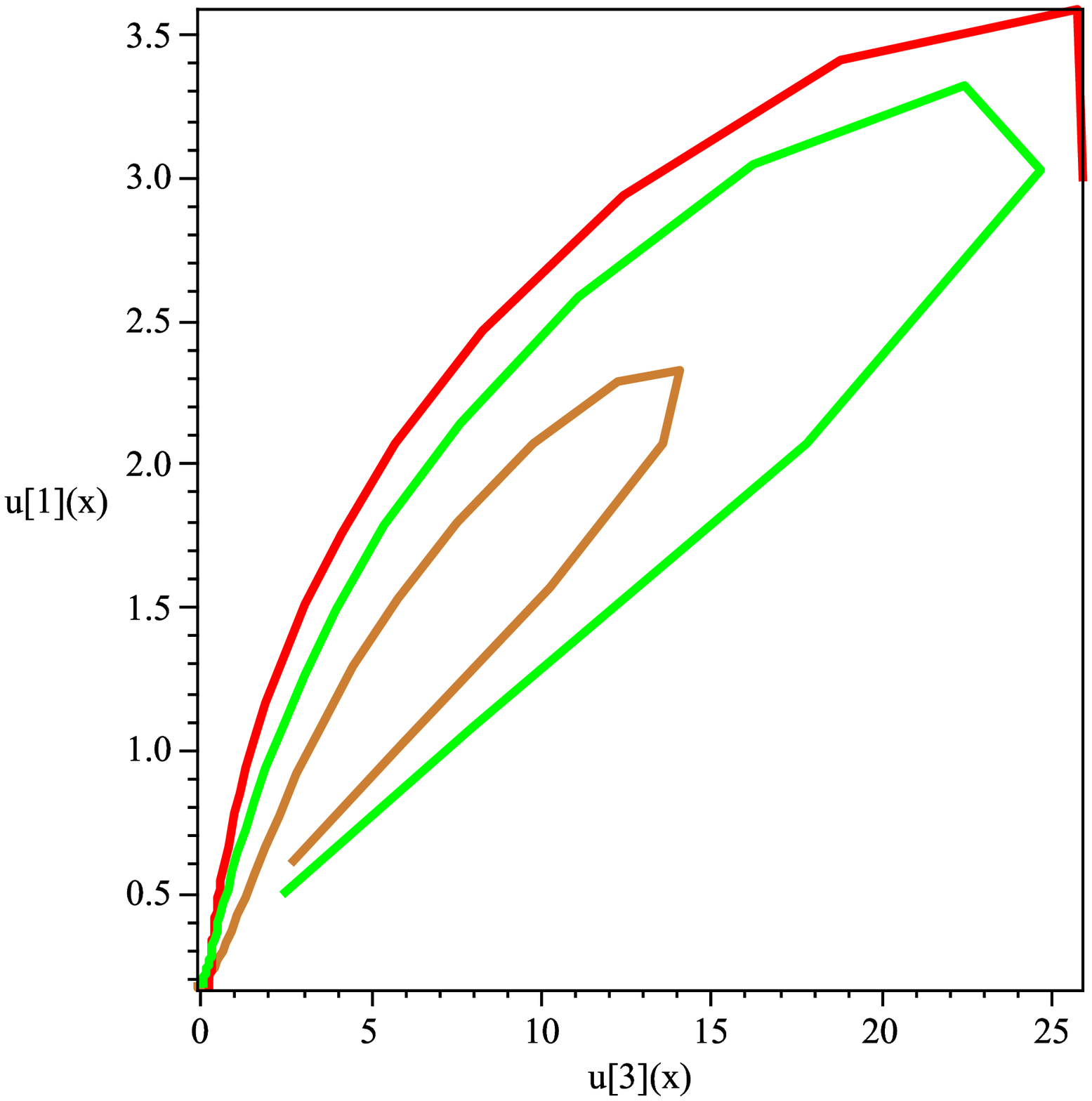}\\
\caption{The phase diagram of the triple interacting fluid model
with the choice of couplings [$\lambda_1=0.2,
\lambda_2=0.4,\lambda_3=0.6$], $x=-1,..,1$ and EoS parameters [
$\omega_1=0,\omega_2=1/3,\omega_3=-1$]. Chosen scene is between
$u_1$ and $u_3$. The curves correspond to the initial conditions as
taken in Fig.1.}
\end{figure}
\newpage
\begin{figure}
\includegraphics[scale=.7]{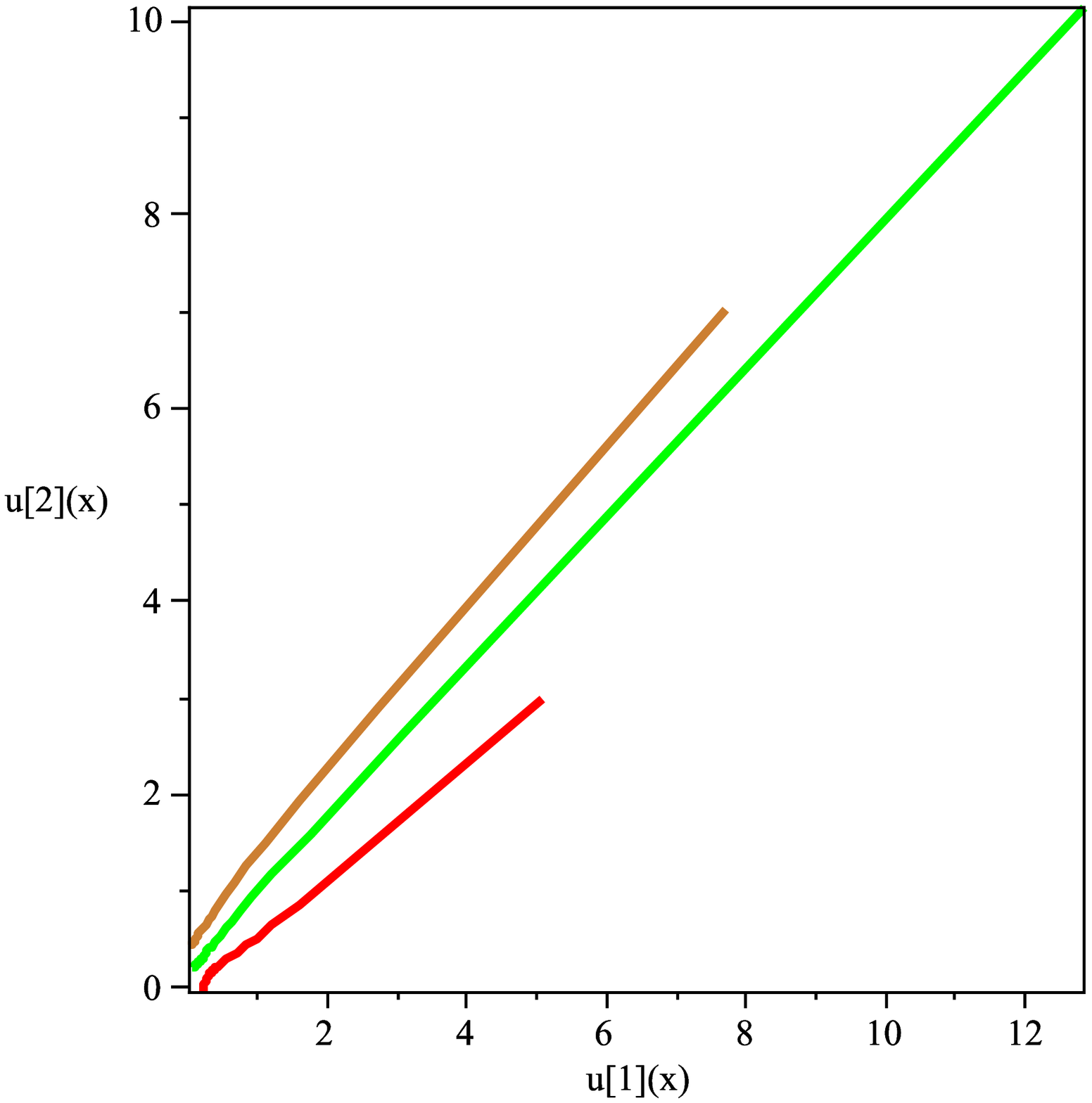}\\
\caption{The phase diagram of the triple interacting fluid model
with the choice of couplings [$\lambda_1=0,
\lambda_2=0.4,\lambda_3=-0.4$], $x=-1,..,1$ and EoS parameters [
$\omega_1=0,\omega_2=1/3,\omega_3=-1/3$]. Chosen scene is between
$u_1$ and $u_2$. The curves correspond to the initial conditions as
taken in Fig.1.}
\end{figure}
\newpage
\begin{figure}
\includegraphics[scale=.7]{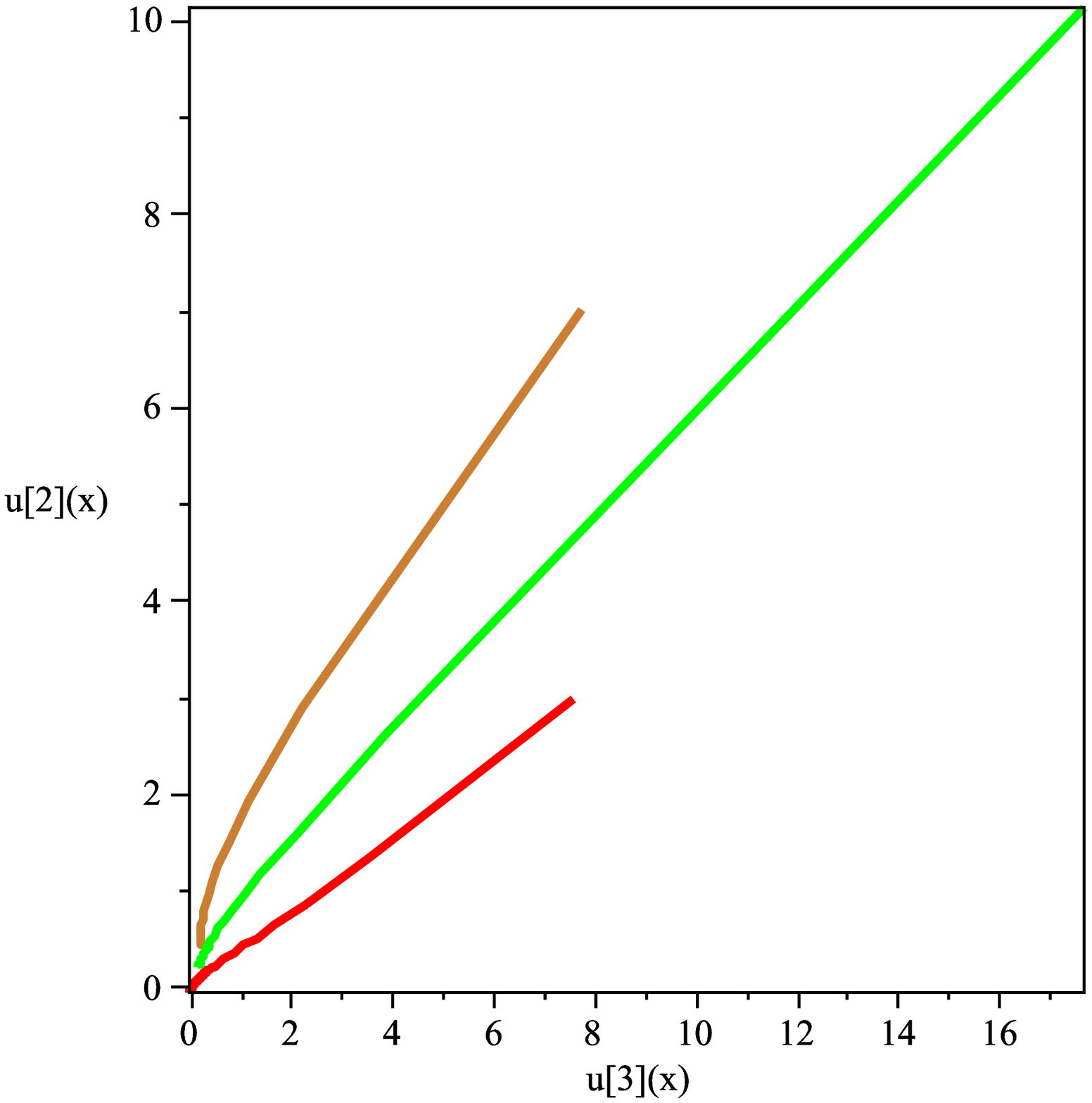}\\
\caption{The phase diagram of the triple interacting fluid model
with the choice of couplings [$\lambda_1=0,
\lambda_2=0.4,\lambda_3=-0.4$], $x=-1,..,1$ and EoS parameters [
$\omega_1=0,\omega_2=1/3,\omega_3=-1/3$]. Chosen scene is between
$u_3$ and $u_2$. The curves correspond to the initial conditions as
taken in Fig.1.}
\end{figure}
\newpage
\begin{figure}
\includegraphics[scale=.7]{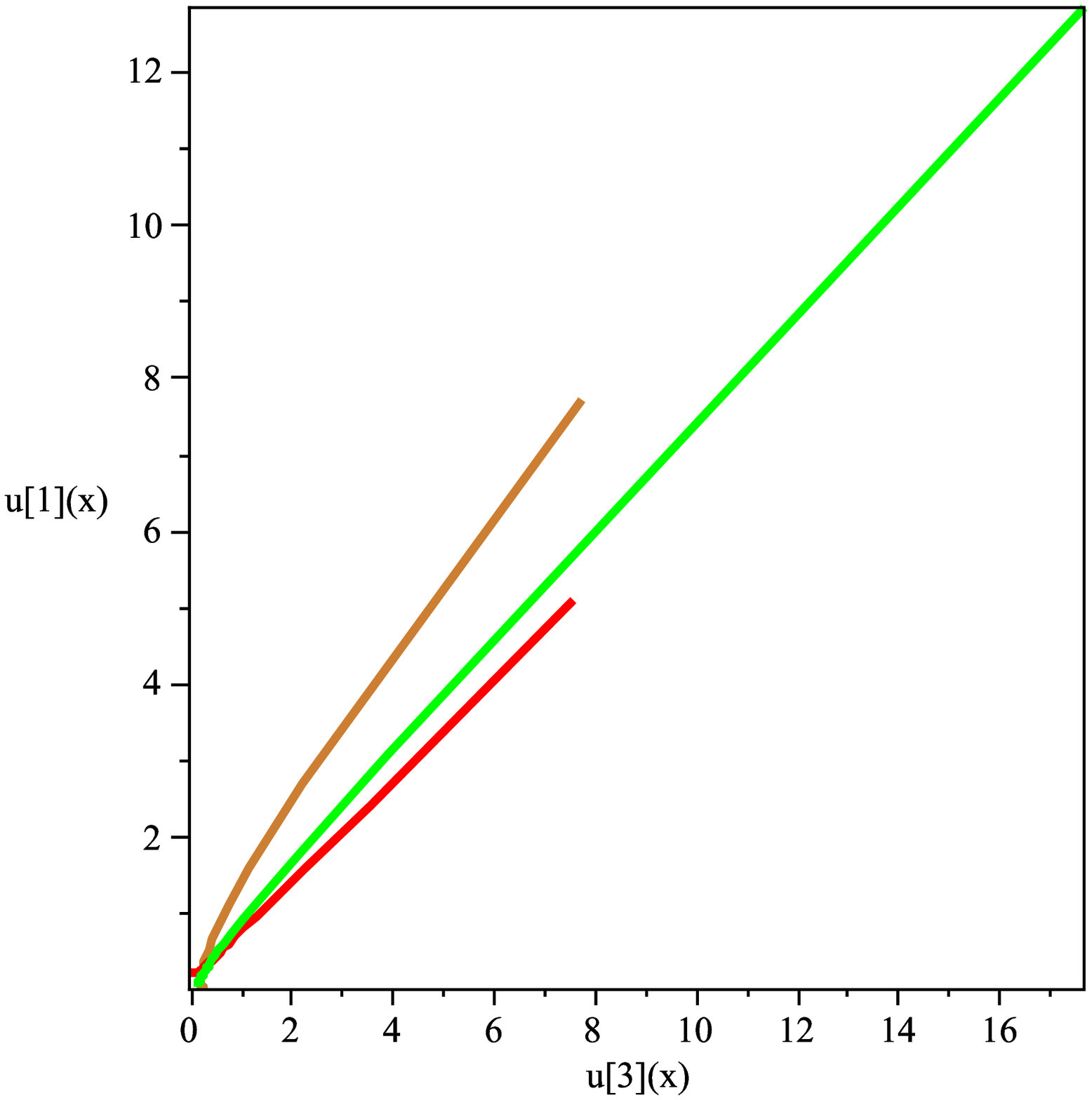}\\
\caption{The phase diagram of the triple interacting fluid model
with the choice of couplings [$\lambda_1=0,
\lambda_2=0.4,\lambda_3=-0.4$], $x=-1,..,1$ and EoS parameters [
$\omega_1=0,\omega_2=1/3,\omega_3=-1/3$]. Chosen scene is between
$u_1$ and $u_3$. The curves correspond to the initial conditions as
taken in Fig.1.}
\end{figure}

\subsubsection*{Acknowledgments}
One of us (MJ) would like to thank John D. Barrow for useful
correspondence. Useful comments from the anonymous referee are also
gratefully acknowledged.
\end{document}